\documentclass{optica-article}

\journal{opticajournal} 

\articletype{Research Article}

\usepackage{amsmath}
\usepackage{bm}
\usepackage{booktabs}
\usepackage{array}
\usepackage{tabularx}

\newcommand{\vp}{\varphi}
\newcommand{\bDelta}{\boldsymbol{\Delta}}
\newcommand{\bM}{\mathbf{M}}
\newcommand{\bT}{\mathbf{T}}
\newcommand{\bP}{\mathbf{P}}
\newcommand{\bR}{\mathbf{R}}

\newcommand{\beps}{\boldsymbol{\varepsilon}}
\newcommand{\bStokes}{\mathbf{S}}

\newcommand{\diag}{\operatorname{diag}}
\newcommand{\Real}{\operatorname{Re}}
\newcommand{\Imag}{\operatorname{Im}}
\newcommand{\nhat}{\hat{n}}
\newcommand{\dhat}{\hat{d}}

\begin{document}

\title{The Electrodynamic Basis of Dichroism-Mediated Polarization Perception}

\author{G.\,P.\ Misson,\authormark{1,*} D.\ Sarenac,\authormark{2} D.\,A.\ Pushin,\authormark{3} and S.\,E.\ Temple\authormark{1,4,5}}

\address{\authormark{1}School of Optometry, Aston University, Birmingham B4 7ET, UK\\
\authormark{2}Department of Physics, University at Buffalo, State University of New York, Buffalo, New York 14260, USA\\
\authormark{3}Institute for Quantum Computing and Department of Physics and Astronomy, University of Waterloo, Waterloo, Ontario N2L 3G1, Canada\\
\authormark{4}Division of Research and Innovation, University of Bristol, Bristol BS8 1QU, UK\\
\authormark{5}Azul Optics Ltd, Henleaze, Bristol BS9 4QG, UK}

\email{\authormark{*}g.misson@aston.ac.uk}

\begin{abstract*}
Humans see the polarization of light through entoptic percepts arising from the macula's Henle fiber layer, where xanthophyll pigments absorb preferentially across the radiating fibers. From the layer's complex dielectric tensor alone, Maxwell's equations in Berreman $4\times4$ form yield its Mueller matrix, set by four scalars $\{A,B,C,D\}$. Its intensity channel depends only on the dichroic pair $A$ and $B$, which fix the percepts at a maximum contrast $|B|/A\approx0.05$. One relation generates the whole dichroism-mediated family: Haidinger's brushes under a uniform field, their dark arms perpendicular to the $\mathbf{E}$-vector; fractured brushes under spatially varying fields; and $N$-fold brushes under vector-vortex illumination.
\end{abstract*}


\section{Introduction}
\label{sec:intro}

Human polarization perception encompasses a hierarchy of visual phenomena
of increasing spatial complexity.  The simplest is the percept evoked by a
spatially uniform linearly polarized field, Haidinger's brushes
\cite{haidinger1844}, a low-contrast yellow-and-blue hourglass figure
that appears about the point of fixation.  More complex percepts arise from spatially
structured polarization patterns, in which the polarization direction varies
across the visual field \cite{misson2015,misson2017}, and from vector-vortex
stimuli carrying orbital angular momentum (OAM), in which the polarization
structure varies with azimuth
\cite{sarenac2020,pushin2026emerging}.  Despite their differing
appearance, these percepts share a single physical origin.  The Henle fiber
layer (HFL) of the macula absorbs blue light more strongly along one
direction than another, and because its fibers radiate from the foveal
center that preferred direction turns with position across the fovea.  This
directional absorption is macular dichroism, and it is carried
mathematically by the imaginary part of the layer's dielectric tensor.  The
retinal response at each point is governed by the same local optical
properties whatever the global structure of the incident field, so the
uniform-field case, Haidinger's brushes, is both the simplest instance and
the necessary foundation for the more general analyses.

Polarization-related visual phenomena can arise from mechanisms other than by dichroic selective absorption.  Boehm's brush
\cite{boehm1940}, and its spin--orbit
coupled light variants \cite{pushin2026boehm}, are driven by angular
variation in scattering strength relative to the local polarization
direction rather than by absorption anisotropy. They are weak at the fovea, where the retina is thinnest, and are most readily seen in the parafoveal field \cite{pushin2026boehm,vos1964}, outside the zone of densest macular pigment. The present paper accounts for the
dichroism-mediated family, and makes no claim on percepts of scattering
origin.

\begin{figure}[htbp]
\centering
\includegraphics[width=\textwidth]{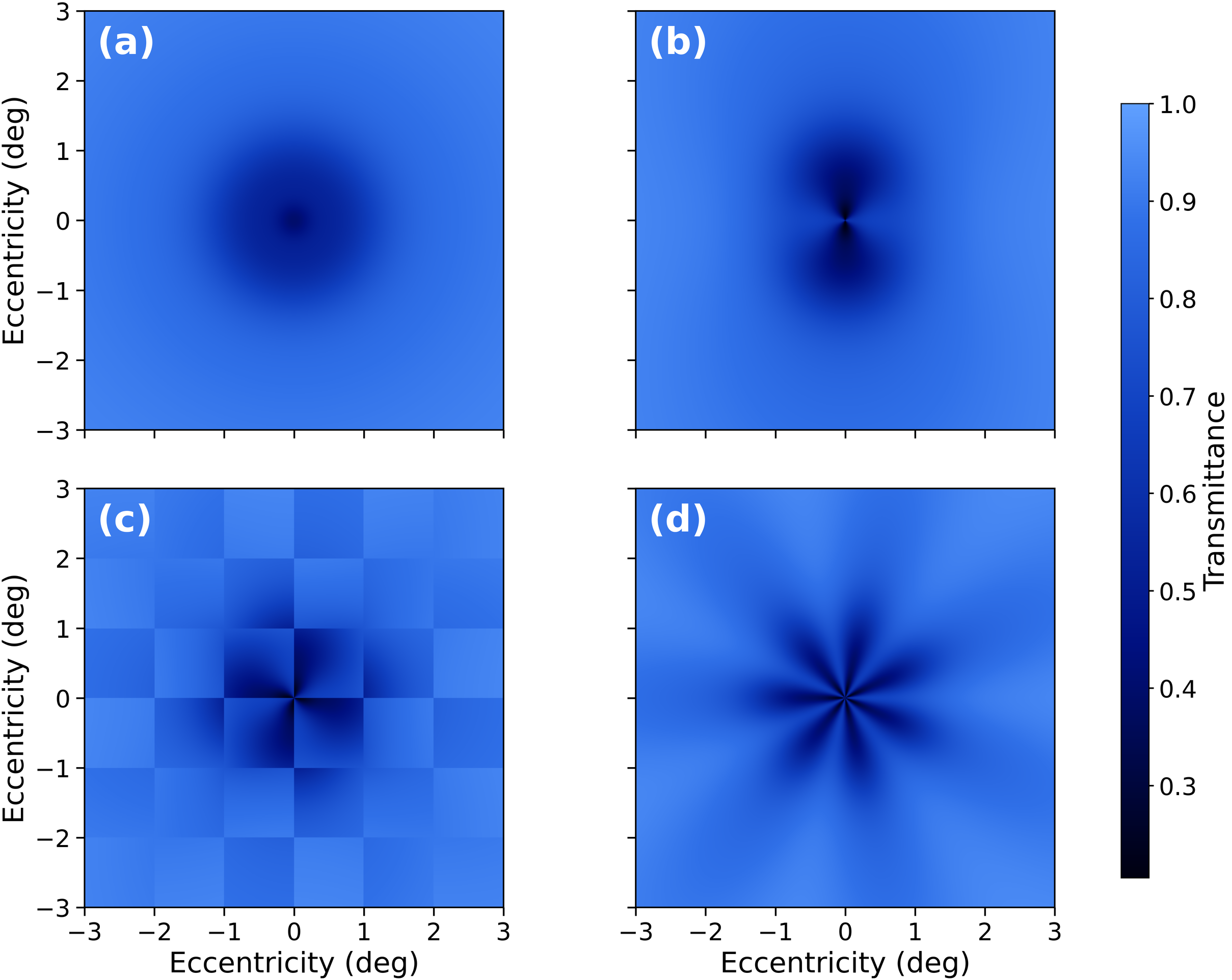}
\caption{%
  One intensity relation, four percepts.  All four panels are computed from
  $I(r,\vp)=A(r)+p\,B(r)\cos[2(\vp-\alpha(r,\vp))]$ with identical optics
  ($\lambda=460\,$nm, $\mathcal{R}=1.115$, $D_0=0.399$, $h=30\,\mu$m,
  corneal retarder $\delta_c=0.08\lambda$) and the
  $\mathrm{M}_{3\mathrm{G}}$ macular pigment profile \cite{misson2025};
  only the incident polarization field differs between
  them.  Each panel covers the central $\pm3^\circ$ of the visual field and
  all four share one color scale.
  (a)~Depolarized input, $p=0$: the dichroic term vanishes identically and
  only the polarization-independent absorption disc of Maxwell's spot
  remains.  Panels (b)--(d) have $p=1$ and differ only in $\alpha$.
  (b)~Uniform linear polarization, $\alpha=0$: Haidinger's brushes, the
  dark arms perpendicular to the horizontal $\mathbf{E}$-vector.
  (c)~A $1^\circ$ checkerboard alternating between vertical and horizontal
  polarization: the brushes fracture into cells of opposite sign.
  (d)~A vector-vortex field of topological charge $\ell=9$,
  $\alpha=\tfrac{1}{2}\ell\vp$, giving $I=A(r)+B(r)\cos[(2-\ell)\vp]$ for
  the macular layer alone and hence $N=|\ell-2|=7$ dark arms.  With the
  corneal retarder included the azimuthal factor also carries a weaker
  $|\ell+2|$ harmonic, of relative amplitude $\tan^2(\delta_c/2)=0.066$
  here; the arm count is unaffected, but the arms are unequal in depth by
  about $12\%$ and unevenly spaced by up to $1^\circ$ of azimuth
  (Supplement~1, Section~S7.5).
  For visibility in print the dichroic coefficient $B$ is scaled tenfold
  ($k_1=0.60$, $k_2=0.20$); because $(k_1+k_2)/2$ coincides with the
  physical mean transmittance $A=0.400$, the scaling alters the modulation
  only and leaves panel~(a), which has no dichroic term, unchanged.  The
  maximum physical contrast, at the foveal center, is $|B|/A=0.05$.}
\label{fig:percepts}
\end{figure}

A companion paper \cite{misson_paper1} derives a unified
complex dielectric tensor for the HFL from molecular biophysics and
effective medium theory, with the real part encoding form birefringence
and the imaginary part encoding macular dichroism.  The present paper asks
what that tensor predicts we should see: first for a uniform field, and
then, because the interaction is local, for any spatially modulated polarization field.  The result, developed here via the Berreman $4\times4$ matrix
formalism, is a Mueller--Stokes model with an intensity channel depending
exclusively on the dichroic (imaginary) component of the tensor.  It
establishes that Haidinger's brushes are a first-order dichroic effect,
and that the remaining dichroism-mediated percepts follow from the same
relation with no further physics (Fig.~\ref{fig:percepts}).

Quantitative models of Haidinger's brushes have adopted phenomenological
optical formalisms.  For example, Mueller calculus was used \cite{misson2003} to represent the
macular layer as an ideal linear radial diattenuator with a corneal retarder,
predicting rotation behavior and contrast variation.  The
quantitative appearance of the brushes was subsequently reproduced by
incorporating
measured dichroic ratios into the model \cite{misson2019}.  The present
paper grounds this phenomenology in electromagnetic first principles by
taking the dielectric tensor as input and deriving the
Mueller matrix from Maxwell's equations without additional assumptions.

The derivation proceeds via the Berreman $4\times4$ matrix formalism
\cite{berreman1972}, which provides an exact treatment of light
propagation in stratified anisotropic media.  The power of this approach
for macular optics lies in three properties: (i)~the isolation of the
single tensor element that mixes the two transverse channels;
(ii)~the factorization of the propagation matrix into a $\vp$-independent
core and a geometric rotation, revealing that all azimuthal dependence is
purely geometric; and (iii)~the natural separation of dichroic ($A,B$) and
retardance ($C,D$) contributions to the Mueller matrix, $\bM(\vp)$.  Thus
within the intensity channel of $\bM(\vp)$, Haidinger's brushes contrast
depends only on $A$ and $B$, with birefringence confined to the retardance
scalars $C$ and $D$.

Following the introduction, Section~\ref{sec:tensor} imports the tensor
and the properties of it that the Berreman analysis requires; Supplement~1
(Sections~S2--S3) develops the Berreman $4\times4$ matrix and its
eigendecomposition.  Sections~\ref{sec:jones}--\ref{sec:mueller} derive
the Jones and Mueller matrices and recover the phenomenological model
\cite{misson2003} as the perfect-dichroism limit.  Section~\ref{sec:discussion}
then lets the polarization azimuth vary with position and recovers the
remaining members of the family.  The corneal-retarder cascade, the
numerical implementation behind Fig.~\ref{fig:percepts}, and the treatment
of individual variation in adaptation are developed in Supplement~1.

\paragraph{Assumptions.}
Light is taken to be monochromatic at $460\,$nm, the macular-pigment
absorption maximum, and normally incident ($\xi=0$ in Berreman's notation
\cite{berreman1972}).  Both restrictions are relaxable within the same
formalism; spectral convolution and oblique-ray $E_z$ coupling are left to
a subsequent paper.


\section{Dielectric tensor and slab model}
\label{sec:tensor}

The geometry, the tensor and its calibration are taken without
modification from the companion paper \cite{misson_paper1}, and are set
out here only as far as the Berreman analysis needs them.  In outline, the
tensor is complex: its real part describes how fast light travels along
and across the fibers, and so the birefringence, while its imaginary part
describes how strongly light is absorbed along each of those two
directions, and so the dichroism.  Both anisotropies are of order
$10^{-3}$ and both are tied to the same fiber axis, which is what makes a
single tensor sufficient.  The frame has
its origin at the foveal center, $z$ anterior along the visual axis so
that light propagates along $-z$, $x$ and $y$ in the retinal plane, and
$\vp$ increasing counterclockwise from $+\hat{x}$ in the clinical fundus
view.  A left eye is assumed throughout, a right eye being
mirror-symmetric.  Radial fibers therefore have direction:


\begin{equation}
  \nhat(\vp) = (\cos\vp,\,\sin\vp,\,0).
  \label{eq:fiber_dir}
\end{equation}
Averaging the xanthophyll absorption over the cylindrical fiber membrane,
which the companion paper sets out in full, leaves an effective absorption
axis lying in the retinal plane at right angles to $\nhat$:
\begin{equation}
  \dhat(\vp) = (-\sin\vp,\,\cos\vp,\,0).
  \label{eq:dipole_dir}
\end{equation}
Absorption is therefore greatest for the electric field
$\mathbf{E}\parallel\dhat$, across the fiber, and least along it.  This is
the geometric content of the Bone model \cite{bone1984}, independently
supported by the molecular-dynamics results of Grudzinski \textit{et al.}
\cite{grudzinski2017}.

The Henle fiber layer enters the
electromagnetic problem as a single slab of thickness $h$: non-magnetic,
optically inactive, and with its entire optical content carried by the
permittivity tensor $\beps(\vp)$ through the constitutive relation
$\mathbf{D}=\beps(\vp)\mathbf{E}$, in which $\mathbf{D}$ is the electric
displacement field and $\mathbf{E}$ the electric field.  Other retinal layers are ignored and the treatment
is foveal, where the Henle fiber layer dominates the polarization
signal.  Referred to its own
principal axes the tensor is diagonal and uniaxial about the fiber:
\begin{equation}
  \beps_{\rm fib}
  = \diag\!\left(\varepsilon_\parallel,\;\varepsilon_\perp,\;\varepsilon_\perp\right),
  \label{eq:eps_fib}
\end{equation}
with complex principal permittivities:
\begin{equation}
  \varepsilon_{\parallel,\perp}
  = \varepsilon_b \pm \tfrac{1}{2}\Delta\varepsilon_r + i\kappa_{\parallel,\perp}.
  \label{eq:complex_eps}
\end{equation}
The cylindrical membrane average makes the out-of-plane element equal to
the in-plane perpendicular element, $\varepsilon_{zz}=\varepsilon_\perp$,
and the off-diagonal elements vanish,
$\varepsilon_{xz}=\varepsilon_{yz}=0$.  At normal incidence the
longitudinal field $E_z$ is not excited, $\varepsilon_{zz}$ drops out of
the transmission problem (Supplement~1, Section~S2), and only the
transverse $2\times2$ block enters the analysis below.

Three constants set the scale.  The background
$\varepsilon_b=n_b^2$ ($n_b\approx1.34$) sets the mean level only.  The
real increment $\pm\tfrac{1}{2}\Delta\varepsilon_r$, with
$\Delta\varepsilon_r=2n_b\Delta n_r>0$, places the slow axis radially.
The imaginary pair $\kappa_{\parallel,\perp}=2n_bk_{\parallel,\perp}$ is
the larger across the fiber.  Of the three, only the imaginary pair
survives into the intensity channel (see Section~\ref{sec:mueller}); this is why the present analysis needs no independent
birefringence calibration.  Its numerical scale follows from two measured
quantities: the dichroic ratio
$\mathcal{R}=\mathrm{OD}_\perp/\mathrm{OD}_\parallel=1.115$, the ratio of
the optical densities across and along the fiber, and the peak macular
pigment optical density at the foveal center, $D_0=0.399$.  At
$\lambda=460\,$nm these give principal transmittances
$T_\parallel=0.42$ and $T_\perp=0.38$ for use in
Sections~\ref{sec:jones}--\ref{sec:mueller}.

Let $R(\vp)$ be the ordinary $3\times3$ rotation of the coordinate axes
through $\vp$ about $\hat{z}$,
\[
  R(\vp) = \begin{pmatrix}
    \cos\vp & -\sin\vp & 0\\
    \sin\vp &  \cos\vp & 0\\
    0        &  0        & 1
  \end{pmatrix},
\]
not to be confused with the $4\times4$ Stokes rotation $\bR(\vp)$ of
Supplement~1, Section~S6, which acts on Stokes vectors and turns through
$2\vp$.  Rotating the fiber frame to azimuth $\vp$,
$\beps_{\rm lab}(\vp)=R(\vp)\,\beps_{\rm fib}\,R(\vp)^\top$, leaves
$\varepsilon_{zz}$ unchanged and gives the transverse block:
\begin{align}
  \varepsilon_{xx}(\vp) &= \varepsilon_{\rm avg}
    + \tfrac{1}{2}\Delta\varepsilon\cos2\vp, \label{eq:exx_lab}\\
  \varepsilon_{yy}(\vp) &= \varepsilon_{\rm avg}
    - \tfrac{1}{2}\Delta\varepsilon\cos2\vp, \label{eq:eyy_lab}\\
  \varepsilon_{xy}(\vp) &= \varepsilon_{yx}(\vp)
    = \tfrac{1}{2}\Delta\varepsilon\sin2\vp, \label{eq:exy_lab}
\end{align}
in which $\varepsilon_{\rm avg}=(\varepsilon_\parallel+\varepsilon_\perp)/2$
and $\Delta\varepsilon=\varepsilon_\parallel-\varepsilon_\perp$.
Although each element varies with $\vp$, the combination
$\varepsilon_{xx}\varepsilon_{yy}-\varepsilon_{xy}^2$ does not, being
equal to $\varepsilon_\parallel\varepsilon_\perp$ at every azimuth.  That invariance
carries through to the Berreman problem, making the eigenvalues of
$\bDelta$ azimuth-free so that the eigenvectors hold all the $\vp$
dependence (Supplement~1, Section~S3).

\begin{figure}[htbp]
\centering
\includegraphics[width=0.86\textwidth]{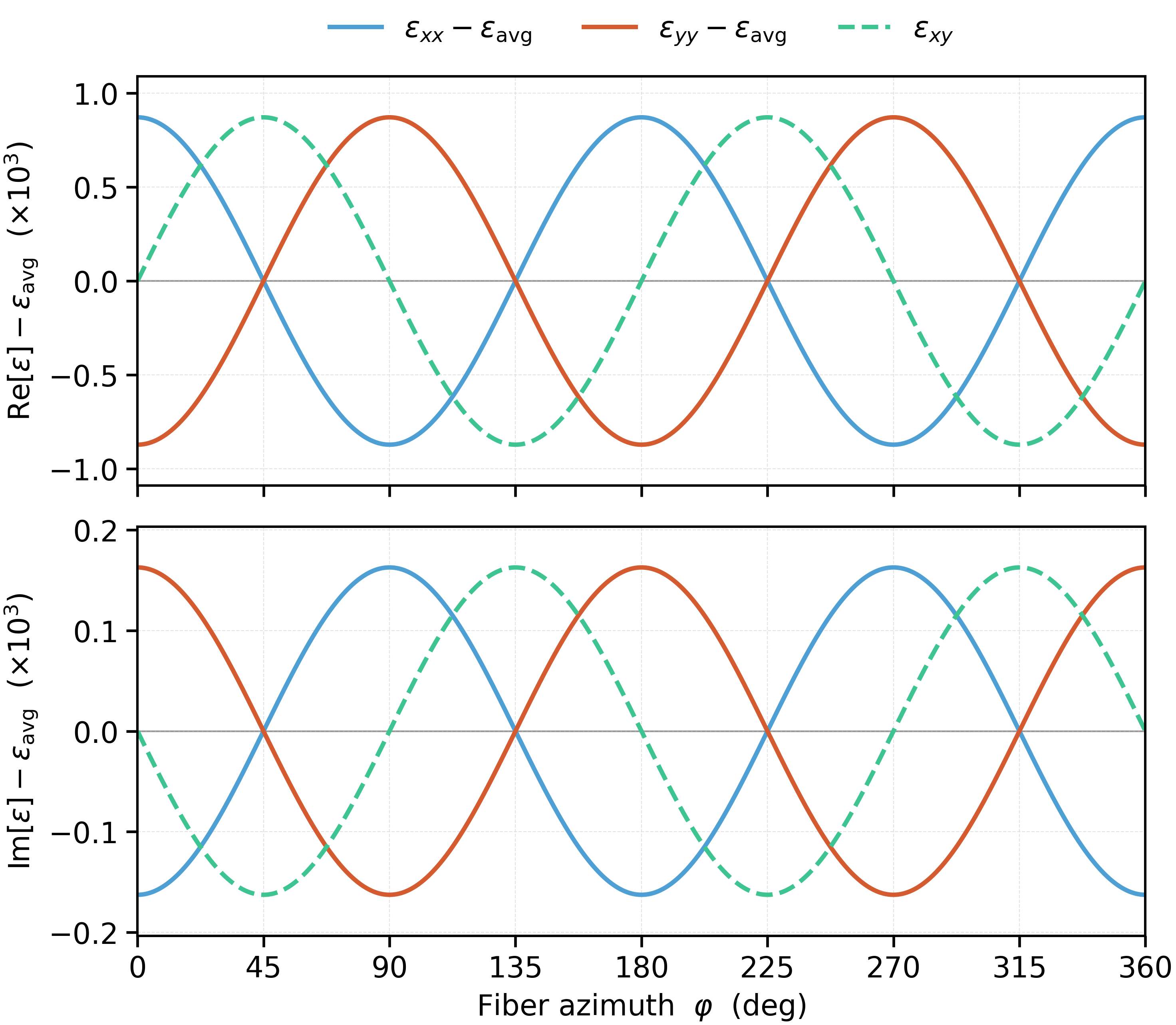}
\caption{%
  Lab-frame dielectric tensor elements of the Henle fiber layer against
  fiber azimuth $\vp$, at $\lambda=460\,$nm with $\mathcal{R}=1.115$,
  $D_0=0.399$, $\Delta n_r=0.65\times10^{-3}$ and $h=30\,\mu$m.  Each
  element is plotted as its departure from the $\vp$-independent mean
  $\varepsilon_{\rm avg}$, scaled by $10^{3}$: real parts (form
  birefringence) above, imaginary parts (dichroism) below.  Note the
  five-fold difference in vertical scale between the frames.  The diagonal
  elements vary as $\pm\cos2\vp$ and the off-diagonal element as
  $\sin2\vp$, in quadrature and, within each frame, of equal amplitude, so
  that $\varepsilon_{xx}\varepsilon_{yy}-\varepsilon_{xy}^2$ is independent
  of $\vp$.  In the lower frame
  $\Imag[\varepsilon_{xx}]<\Imag[\varepsilon_{yy}]$ at $\vp=0$, the
  signature of the perpendicular-dipole orientation that gives $B>0$.}
\label{fig:tensor}
\end{figure}

Figure~\ref{fig:tensor} shows these three elements over a full turn of the
azimuth.  Because they sit on a large $\vp$-independent pedestal
($\varepsilon_{\rm avg}=1.7956+3.00\times10^{-3}i$), each is plotted as its
departure from that pedestal, scaled by $10^{3}$, with the real parts above
and the imaginary parts below.  In each frame the two diagonal elements
vary as $\pm\cos2\vp$ in antiphase while the off-diagonal element varies as
$\sin2\vp$, in quadrature with them and of equal amplitude:
$|\Real[\Delta\varepsilon]|/2=0.871\times10^{-3}$ in the upper frame and
$|\Imag[\Delta\varepsilon]|/2=0.163\times10^{-3}$ in the lower.  Equality of
amplitude within a frame is precisely the condition that keeps
$\varepsilon_{xx}\varepsilon_{yy}-\varepsilon_{xy}^2$ constant, so
Fig.~\ref{fig:tensor} is a direct visualization of that invariance:
everything varying with azimuth is a rotation of a fixed pair of principal
axes, not a change in the medium.

The two frames differ in scale by a factor of $5.4$.  The form
birefringence is the larger anisotropy, yet it is confined to the
retardance scalars $C$ and $D$ and absent from the intensity channel. It is therefore the weaker imaginary modulation that the observer sees.  At $\vp=0$ the
lower frame places $\Imag[\varepsilon_{xx}]=2.84\times10^{-3}$ below
$\Imag[\varepsilon_{yy}]=3.17\times10^{-3}$: a parallel-dipole model would
interchange those two curves and leave the upper frame untouched.

\paragraph{Sign of $\Delta\varepsilon$.}
Because $\kappa_\perp>\kappa_\parallel$, we have
$\Imag[\Delta\varepsilon]=\kappa_\parallel-\kappa_\perp<0$ giving
$\Imag[\varepsilon_{xx}]<\Imag[\varepsilon_{yy}]$ at $\vp=0$.  Absorption
at that azimuth is therefore weakest along the fiber and strongest across
it, which is the perpendicular-dipole orientation.

The real part of the same tensor governs the form birefringence
($\Delta n_r\approx0.65\times10^{-3}$, slow axis radial).  It reaches the
Mueller matrix through the retardance scalars $C$ and $D$ and leaves the
transmitted intensity untouched (Section~\ref{sec:mueller}).  Its own
observable is the macular cross, which is left to further study.

Carried through the radial arrangement of the Henle fibers about the
fovea, this single tensor is the common origin of the macular polarization
percepts (Fig.~\ref{fig:percepts}), as the Mueller-matrix derivation
below makes explicit.

\section{Jones transmission matrix and the angular intensity relation}
\label{sec:jones}

The key results of the Berreman treatment are summarized here; the full
development of the $4\times4$ matrix, its eigendecomposition and the
resulting propagation matrix is given in Supplement~1 (Sections~S2--S3).
Starting from Maxwell's equations in the $6\times6$ form of Berreman
\cite{berreman1972}, the $4\times4$ propagation matrix at normal incidence
is:
\begin{equation}
  \bDelta(\vp) = \begin{pmatrix}
    0 & 1 & 0 & 0 \\
    \varepsilon_{xx}(\vp) & 0 & \varepsilon_{xy}(\vp) & 0 \\
    0 & 0 & 0 & 1 \\
    \varepsilon_{xy}(\vp) & 0 & \varepsilon_{yy}(\vp) & 0
  \end{pmatrix},
  \label{eq:Delta}
\end{equation}
which is block-diagonal only where $\varepsilon_{xy}$ vanishes at
$\vp=0$ and $\vp=90^\circ$ and every $180^\circ$ thereafter.  Its
eigenvalues, $q_{1,2}=\pm\sqrt{\varepsilon_\parallel}$ and
$q_{3,4}=\pm\sqrt{\varepsilon_\perp}$, carry no $\vp$ dependence at all, so
the fiber orientation is encoded entirely in the eigenvectors, equivalently
in the $\vp$-dependent combination of $\varepsilon_{xx}-\varepsilon_{yy}$
and $\varepsilon_{xy}$; $\varepsilon_{xy}$ describes the mixing of the two
laboratory transverse channels when the principal axes are not aligned with
them.  All
azimuthal information therefore resides in the eigenvectors, and the
propagation matrix factorizes as
$\bP(h,\vp)=\mathbf{V}(\vp)\,\bP_0(h)\,\mathbf{V}(-\vp)$, in which
$\mathbf{V}(\vp)$ is the field-space rotation through the fiber azimuth:
a principal-axis propagator viewed through a geometric rotation
(Supplement~1, Section~S3).

Under the matched-index boundary approximation (the layer interfaces are
near index-matched to the surrounding ocular media, so boundary
reflections are negligible), the resulting Jones matrix is:
\begin{equation}
  \bT(\vp) = \frac{e^{i\delta_1}+e^{i\delta_2}}{2}\,\mathbf{I}_2
  + \frac{e^{i\delta_1}-e^{i\delta_2}}{2}
  \begin{pmatrix}\cos2\vp & \sin2\vp\\\sin2\vp & -\cos2\vp\end{pmatrix},
  \label{eq:jones}
\end{equation}
and for incident light that is fully linearly polarized at azimuth
$\alpha$, so that the incident Jones vector is
$\mathbf{e}_\alpha=(\cos\alpha,\sin\alpha)^\top$, the transmitted
intensity is:
\begin{equation}
  I(\vp,\alpha) = |\bT\mathbf{e}_\alpha|^2
  = \underbrace{\tfrac{T_\parallel+T_\perp}{2}}_{A}
  + \underbrace{\tfrac{T_\parallel-T_\perp}{2}}_{B}\cos[2(\vp-\alpha)],
  \label{eq:intensity}
\end{equation}
where $T_{\parallel,\perp}=\exp[-2\,\Imag(\delta_{1,2})]$ are the
principal intensity transmittances.  Equation~(\ref{eq:intensity}) is a
single-point, fully-polarized result: it is stated at one retinal location,
so that $A$ and $B$ carry no argument, and it presumes a definite incident
azimuth $\alpha$.  Both restrictions are lifted in
Section~\ref{sec:mueller}, where the Stokes form
(Eq.~\ref{eq:S0out}) admits arbitrary incident polarization, and in
Section~\ref{sec:discussion}, where the macular pigment profile makes $A$
and $B$ functions of eccentricity.  With the perpendicular-dipole
orientation $T_\parallel>T_\perp$, so $B>0$.  The $\cos[2(\vp-\alpha)]$
dependence makes the transmittance a two-fold function (period
$180^\circ$) that rotates rigidly with the polarization direction
$\alpha$.  While the transmittance has two bright and two dark lobes, the
perceived brushes show only two dark arms as a consequence of neural
adaptation \cite{misson2020}.  The polarization-independent part of the
macular-pigment absorption, the radially symmetric Maxwell's-spot disc, is
a stationary retinal image that is progressively cancelled by Troxler
fading during steady fixation.  Once this uniform background is
discounted, only the two absorption maxima survive as a departure from it,
giving two-armed dark brushes perpendicular to $\mathbf{E}$; the light
brushes are simply the adapted background seen where absorption is least.
Figure~\ref{fig:percepts}(a,b) shows the two components separately.  The
full perceptual model, including the variation between observers in how
completely that background is discounted and hence whether the light
brushes are seen as distinctly brighter, is developed in Supplement~1
(Section~S7.6).

\section{Mueller matrix formulation and the perfect-dichroism limit}
\label{sec:mueller}

The $4\times4$ Mueller matrix is obtained via:
\begin{equation}
  \bM = \mathbf{U}\,(\bT\otimes\bT^*)\,\mathbf{U}^{-1},\qquad
  \mathbf{U} = \frac{1}{\sqrt{2}}
  \begin{pmatrix}1&0&0&1\\1&0&0&-1\\0&1&1&0\\0&i&-i&0\end{pmatrix}.
  \label{eq:J2M}
\end{equation}

The Mueller matrix is fully determined by four real scalars:
\begin{align}
  A &= \tfrac{1}{2}(T_\parallel+T_\perp),
     && \text{(mean transmittance)} \label{eq:A}\\
  B &= \tfrac{1}{2}(T_\parallel-T_\perp)>0,
     && \text{(dichroic coefficient)} \label{eq:B}\\
  C &= \sqrt{T_\parallel T_\perp}\,\cos\Delta\phi,
     && \text{(retardance, cosine)} \label{eq:C}\\
  D &= \sqrt{T_\parallel T_\perp}\,\sin\Delta\phi,
     && \text{(retardance, sine)} \label{eq:D}
\end{align}
where $\Delta\phi=\Real[\delta_1]-\Real[\delta_2]=2\pi h\,\Delta n_r/\lambda$
is the phase retardance due to form birefringence.  The maximum
Haidinger's brushes contrast, attained at the foveal center where the
pigment density is greatest, is $|B|/A\approx0.05$ (Eq.~\ref{eq:DoLP}).

Two of these scalars describe absorption and two describe retardance: $A$
and $B$ fix the mean transmittance and the depth of its azimuthal
modulation, while $C$ and $D$ fix the phase relationship the layer imposes
between the two polarization channels.  Their physical roles and their
dependence on the input parameters are collected in
Table~\ref{tab:ABCD}.

\begin{table}[htbp]
\centering
\caption{The four Mueller matrix scalars $\{A,B,C,D\}$ and their physical
  roles.  $T_{\parallel,\perp}=\exp[-2\,\Imag(\delta_{1,2})]$ are the
  principal intensity transmittances ($\Delta\phi$ is the birefringent
  phase retardance).  Numerical values are computed at $\lambda=460\,$nm
  and $h=30\,\mu$m.  Derivations are given in Supplement~1.}
\label{tab:ABCD}
\footnotesize
\renewcommand{\arraystretch}{1.15}
\begin{tabularx}{\textwidth}{%
  >{\centering\arraybackslash}p{1.0cm}
  >{\raggedright\arraybackslash}p{3.2cm}
  >{\raggedright\arraybackslash}p{3.1cm}
  >{\centering\arraybackslash}p{1.5cm}
  >{\raggedright\arraybackslash}X}
\toprule
Scalar & Definition & Physical role & Value & Depends on \\
\midrule
$A$ & $\tfrac{1}{2}(T_\parallel+T_\perp)$ &
  Mean transmittance; sets overall intensity level &
  $0.400$ &
  $\mathcal{R}$, $D_0$ \\
$B$ & $\tfrac{1}{2}(T_\parallel-T_\perp)>0$ &
  Dichroic coefficient; amplitude of the modulation of the brushes, of
  contrast $|B|/A$ &
  $+0.020$ &
  $\mathcal{R}$, $D_0$ \\
$C$ & $\sqrt{T_\parallel T_\perp}\cos\Delta\phi$ &
  Retardance cosine; $S_2$--$S_3$ coupling &
  $0.385$ &
  $\Delta n_r$, $h$, $\lambda$; $\mathcal{R}$, $D_0$ via
  $\sqrt{T_\parallel T_\perp}$ \\
$D$ & $\sqrt{T_\parallel T_\perp}\sin\Delta\phi$ &
  Retardance sine; generates transmitted ellipticity
  $S_3^{\rm out}\propto D\sin[2(\vp-\alpha)]$ &
  $0.105$ &
  $\Delta n_r$, $h$, $\lambda$; $\mathcal{R}$, $D_0$ via
  $\sqrt{T_\parallel T_\perp}$ \\
\midrule
\multicolumn{5}{>{\raggedright\arraybackslash}p{\dimexpr\textwidth-2\tabcolsep\relax}}{Key
  structural property: Row~0 of $\bM(\vp)$ (Eq.~\ref{eq:M_phi}) contains
  only $A$ and $B$; $C$ and $D$ do not enter the intensity channel.} \\
\bottomrule
\end{tabularx}
\end{table}

At $\vp=0$ the Mueller matrix takes the block form:
\begin{equation}
  \bM_0 = \begin{pmatrix}
    A & B & 0 & 0\\
    B & A & 0 & 0\\
    0 & 0 & C & D\\
    0 & 0 & -D & C
  \end{pmatrix}.
  \label{eq:M0}
\end{equation}
The upper-left $2\times2$ block is a linear dichroic attenuator
($S_0$--$S_1$ coupling from the absorption anisotropy); the lower-right
is a linear retarder ($S_2$--$S_3$ coupling from the birefringence).
The block structure confirms the separation of dichroic and birefringent
effects and their independence at the matrix level.

Writing $c=\cos2\vp$, $s=\sin2\vp$:
\begin{equation}
  \bM(\vp) = \begin{pmatrix}
    A     & Bc        & Bs        & 0   \\
    Bc    & Ac^2+Cs^2 & (A-C)sc   & -Ds \\
    Bs    & (A-C)sc   & As^2+Cc^2 & Dc  \\
    0     & Ds        & -Dc       & C
  \end{pmatrix}.
  \label{eq:M_phi}
\end{equation}
Row~0 (transmitted intensity) contains only $A$ and $B$, never the
retardance scalars $C$ or $D$.  This is an exact structural property of
$\bM(\vp)$, not a small-anisotropy approximation.  Column~0 shows that even unpolarized
input acquires partial linear polarization $|B|/A$ after transmission.

Row~0 of Eq.~(\ref{eq:M_phi}) is the general statement of the intensity
relation, and it is worth reading directly:
$S_0^{\rm out}=A\,S_0+B(S_1\cos2\vp+S_2\sin2\vp)$ for \emph{any} incident
Stokes vector.  Writing a general input as
$\bStokes_{\rm in}=(1,\;p\cos2\alpha,\;p\sin2\alpha,\;S_3)^\top$, in which
$p=\sqrt{S_1^2+S_2^2}$ is the degree of \emph{linear} polarization and
$\alpha$ the azimuth of that linear component,
\begin{align}
  S_0^{\rm out} &= A+p\,B\cos[2(\vp-\alpha)], \label{eq:S0out}\\
  S_3^{\rm out} &= C\,S_3+p\,D\sin[2(\vp-\alpha)]. \label{eq:S3out}
\end{align}
$S_0^{\rm out}$ carries the transmitted intensity, and with it the
brushes; $S_3^{\rm out}$ carries the ellipticity generated by the
birefringence.  Three consequences follow, and they fix the scope of every
statement made below.  First, the zero in the fourth entry of Row~0 means
that the circular component $S_3$ never reaches the intensity channel:
circularly polarized light is, for this layer, indistinguishable from
unpolarized light.  Second, elliptical input acts only through its linear
part, at strength $p$ and azimuth $\alpha$.  Third, the depolarized case is
not a separate physical mechanism but the limit $p=0$, at which the
dichroic term vanishes identically and $S_0^{\rm out}=A$: this is
Maxwell's spot, and it is the polarization-\emph{independent} member of the
family.  Setting $p=1$ recovers Eq.~(\ref{eq:intensity}).

Unpolarized input acquires, on transmission, a degree of
linear polarization:
\begin{equation}
  \mathrm{DoLP}_{\rm induced} = \frac{|B|}{A}
  = \frac{T_\parallel-T_\perp}{T_\parallel+T_\perp}
  = \tanh\!\left[\tfrac{1}{2}\ln\!10\,
      \left(\mathrm{OD}_\perp-\mathrm{OD}_\parallel\right)\right]
  \simeq \ln\!10\;D_0\,\frac{\mathcal{R}-1}{\mathcal{R}+1},
  \label{eq:DoLP}
\end{equation}
the last form holding to first order in the optical-density difference.
Substituting the small-$f$ molecular result
$\mathcal{R}\simeq1+\tfrac{3}{2}f\delta_{\rm op}$, in which $f$ is the oriented fraction of the pigment
molecules and $\delta_{\rm op}$ their orientational order parameter \cite{misson_paper1}, gives:
\begin{equation}
  \frac{|B|}{A} \;\simeq\; \tfrac{3}{4}\,\ln\!10\;D_0\,f\,\delta_{\rm op}.
  \label{eq:DoLP_mol}
\end{equation}
Haidinger's brushes contrast is therefore set jointly by how well the
pigment is ordered and by how much of it is present: it scales with $D_0$,
and vanishes together with the pigment.  At $D_0=0.399$ the prefactor
$\tfrac{3}{4}D_0\ln\!10\approx0.69$, and the measured-parameter value,
from $T_\parallel=0.42$ and $T_\perp=0.38$, is
$\mathrm{DoLP}_{\rm induced}\approx0.05$.

\paragraph{The perfect-dichroism limit and the phenomenological model.}
It is worth asking what the layer would look like if it were a perfect
polarizer, absorbing everything across the fibers and nothing along them.
Earlier phenomenological models \cite{misson2003} assumed exactly that, and the present
result reduces to them in that limit.  The limit is
$T_\perp/T_\parallel\to0$, equivalently
$\mathrm{OD}_\perp-\mathrm{OD}_\parallel\to\infty$; normalizing out the
common-mode absorption ($T_\parallel\to1$) then gives
$A\to\tfrac{1}{2}$, $B\to+\tfrac{1}{2}$, $C\to0$, $D\to0$, and
\begin{equation}
  \bM(\vp)\big|_{T_\perp/T_\parallel\to0} = \mathbf{Mp}(a_m),
  \qquad a_m=\vp,
  \label{eq:misson_limit}
\end{equation}
which is the ideal linear-polarizer Mueller matrix \cite{misson2003} of the radial
diattenuator with transmission axis $a_m=\vp$
(the fiber direction).  For fully linearly polarized input ($p=1$),
Eq.~(\ref{eq:S0out}) reduces to
$I=\cos^2(\vp-\alpha)$ (Malus' law), and the dark arms of the
brushes (yellow arms with white polarized light) lie perpendicular to the $\mathbf{E}$-vector, where the fibers run across it
and absorption is greatest. The limit demands more than perfect molecular ordering.  At fixed pigment density,
$\mathcal{R}\to\infty$ drives $\mathrm{OD}_\parallel\to0$ and
$\mathrm{OD}_\perp\to2D_0$, so that $T_\parallel\to1$ but
$T_\perp\to10^{-2D_0}$, which is $0.16$ at the measured $D_0=0.399$ rather
than zero.  Equation~(\ref{eq:DoLP}) then caps the contrast at
$\tanh(D_0\ln\!10)=0.73$.  A perfect polarizer therefore requires
$\mathcal{R}\to\infty$ and  i.e. the available optical
density bounds the contrast however well the pigment is aligned.  The
measured $|B|/A\approx0.05$ is thus some twenty-fold below the ideal
contrast of unity, and some fifteen-fold below the $0.73$ that the foveal
pigment density alone would permit.  The brushes are therefore faint on two counts:
imperfect molecular alignment and limited pigment density.


\section{Discussion}
\label{sec:discussion}

The central result of this paper is that a single complex dielectric
tensor, describing how fast and how strongly the Henle fiber layer
transmits light along and across its fibers \cite{misson_paper1}, taken as
the sole physical input, generates through Maxwell's equations, and
without additional assumptions, a Mueller matrix whose intensity channel
depends only on the dichroic sector $A$ and $B$. The form
birefringence of the Henle fiber layer is separately confined to the polarization-state scalars $C$ and
$D$.  This separation emerges from the
block structure of the Berreman propagation matrix.  Within the intensity
channel (Row~0 of Eq.~\ref{eq:M_phi}) the confinement of the Haidinger's
brushes contrast to the dichroic pair $A$ and $B$ is an exact structural
property of
$\bM(\vp)$, not a leading-order approximation in the anisotropies.  The
birefringence is not quite absent from $(A,B)$ themselves, since
$\Delta\varepsilon_r$ shifts $\sqrt{\varepsilon_{\parallel,\perp}}$ and
hence $T_{\parallel,\perp}$; for the values adopted here, however, that
shift changes $|B|/A$ by under $0.5\%$ of its own value, far below the
uncertainty on any input parameter.  The
Berreman treatment further shows that the eigenvalues of $\bDelta$ are
independent of $\vp$, so that all azimuthal variation is geometric and the
macular layer behaves as a fixed principal-axis medium viewed through a
rotation (Supplement~1, Section~S3).  The sign $B>0$ is the direct,
observable signature of the perpendicular-dipole geometry: the two dark
arms of the brushes lie perpendicular to the $\mathbf{E}$-vector, whereas a
parallel-dipole
model would give $B<0$ and dark arms parallel to it, which is not
observed.

The phenomenological Mueller-matrix model \cite{misson2003} is
recovered exactly as the perfect-dichroism limit
$T_\perp/T_\parallel\to0$
with common-mode absorption removed (Eq.~\ref{eq:misson_limit}).  The
present derivation, therefore, supplies the electromagnetic basis for a
model previously introduced on phenomenological grounds, and replaces its
implicit perfect-polarizer assumption ($B=A=\tfrac12$) with an explicit
contrast $|B|/A\approx0.05$ computed from measured molecular parameters,
some twenty-fold smaller and quantitatively consistent with the low
perceived contrast of the brushes.  Its previously published
elaborations, the corneal-retarder cascade \cite{misson2003}, the
realistic macular-pigment distribution \cite{misson2019,misson2025}, and
the Maxwell's-spot and Haidinger's-brushes partition \cite{misson2020},
are developed in Supplement~1.

Because the transmitted intensity at each retinal point is fixed by the
local dichroic interaction, the same Mueller--Stokes machinery applies
pointwise to any incident polarization field.  Two generalizations of
Eq.~(\ref{eq:intensity}) are needed, and they are independent of one
another.  The incident field may vary with position, so that the local
degree of linear polarization and its azimuth become fields,
$p\to p(r,\vp)$ and $\alpha\to\alpha(r,\vp)$; and the macular pigment
density falls with eccentricity, so that the dichroic scalars become
functions of radius, $A\to A(r)$ and $B\to B(r)$.  The relation used
throughout the remainder of this paper and in Supplement~1 is therefore
\begin{equation}
  I(r,\vp) = A(r) + p(r,\vp)\,B(r)\,\cos\!\bigl[2\bigl(\vp-\alpha(r,\vp)\bigr)\bigr],
  \label{eq:I_general}
\end{equation}
which is Eq.~(\ref{eq:S0out}) with the pigment profile restored; the
construction of $A(r)$ and $B(r)$ from the macular pigment optical density
is given in Supplement~1, Section~S7.1. Figure~\ref{fig:percepts} makes the point directly: the four panels differ
only in the incident polarization field $\{p,\alpha\}(r,\vp)$, the optics
being identical throughout.  Panel~(a) is the depolarized limit $p=0$, at
which the dichroic term vanishes and only the radially symmetric macular pigment absorption
disc (perceived entoptically as Maxwell's spot) remains; the remaining three have $p=1$ and differ
only in $\alpha$.  A uniform field, $\alpha$ constant, gives Haidinger's
brushes (b); a checkerboard of vertical and horizontal polarization
fractures the brushes into cells that alternate in sign (c)
\cite{misson2015,misson2017,temple2015}; and a vector-vortex field, whose
polarization azimuth advances as $\alpha=\tfrac{1}{2}\ell\vp$, gives
$I=A(r)+B(r)\cos[(2-\ell)\vp]$, an $N$-fold figure with $N=|\ell-2|$
\cite{sarenac2020,pushin2026emerging}; panel~(d) shows $\ell=9$, hence
$N=7$.  


Corneal retardation modifies the polarization-dependent percepts.  A corneal retarder of
retardance $\delta_c$ with its fast axis taken along $\hat{x}$  compresses the Stokes $S_2$ axis,
$S_2\to S_2\cos\delta_c$, which reduces the modulation and rotates the
figure by amounts set by the incident azimuth alone (Supplement~1,
Section~S7.4).  Both effects vanish at $\alpha=0$ and $90^\circ$, and both
are largest near $45^\circ$, where $\delta_c=28.8^\circ$ costs $12\%$ of
the modulation.  For a vector vortex $\alpha$ advances with $\vp$, so the
compression varies around the field and the modulation splits exactly into
$\cos^2(\delta_c/2)\cos[(\ell-2)\vp]+\sin^2(\delta_c/2)\cos[(\ell+2)\vp]$
(Supplement~1, Section~S7.5).  The arm count $N=|\ell-2|$ survives while
$\tan^2(\delta_c/2)<(\ell-2)/(\ell+2)$, for $\ell=9$ up to
$\delta_c=77^\circ$ against a physiological $11^\circ$--$43^\circ$
\cite{knighton2002}, though the seven arms then differ in intensity by up to
$12\%$.  The order $N$ is therefore robust to, but not independent of, a
realistic cornea.  Subject to that qualification it is a specific and
testable prediction: the macula is sensitive to the polarization winding
associated with this class of vector-vortex states, and this forward model
provides the electromagnetic foundation for further study.

One limitation of the quantitative contrast deserves emphasis.  The value
$|B|/A\approx0.05$ is obtained by applying the measured bulk dichroic ratio
$\mathcal{R}$ to the peak macular pigment optical density $D_0$, both taken
at the foveal center.  That combination is an upper bound rather than a
prediction of where the percept is strongest, because the two quantities
are not carried by the same pigment pool.  Only the fraction of the
xanthophyll that is oriented, by being bound within the radially running
Henle fibers, contributes to $B$; the unoriented remainder contributes to
$A$ alone.  At the foveal center the Henle fibers are absent, so the
oriented fraction, and with it the polarization-dependent signal, must fall
to zero however dense the pigment is there.  The three-Gaussian model used
here \cite{misson2025} shows this in its own decomposition: its central
zeaxanthin component peaks at $r=0$, whereas its Henle component peaks at
$r=0.56^\circ$ and retains only $58\%$ of its own maximum at the center,
where it accounts for under a quarter of the total density.  A
contrast tied to the oriented pool alone would therefore be maximal in an
annulus rather than at the center, as the circularly-oriented macular
pigment measurements of Pushin \textit{et al.} \cite{pushin2025compod}
indicate.  Resolving this requires an eccentricity-resolved oriented
fraction $f(r)$, which the present tensor does not supply; the simulations
here follow the published two-dimensional treatment in scaling both $A$ and
$B$ by the total pigment profile \cite{misson2019,misson2020}, and the
quoted $0.05$ should be read as the maximum contrast the measured
parameters permit, not as the contrast at any particular eccentricity.

The treatment is monochromatic ($\lambda=460\,$nm) and restricted to
normal incidence; white-light viewing requires spectral convolution with
the macular-pigment absorption and cone sensitivities (which sets the
yellow-blue chromatic appearance), and oblique rays introduce small $E_z$
cross-terms at larger eccentricities.  The optics are treated as
coherent, so incoherent ocular scatter enters only as a
polarization-insensitive background that lowers contrast without altering
the angular symmetry, and the adaptation model is deliberately minimal,
capturing the cancellation of the polarization-independent background, and
its variation between observers (Supplement~1, Section~S7.6), but not the
temporal dynamics or cortical filtering of the percept.  None of these affects the principal result, which is an
algebraic property of the Mueller matrix.  Clinically, impaired
Haidinger's brushes perception in macular disease \cite{muller2016} and
polarization-pattern diagnostics \cite{misson2020tvst,misson2024} may be
related through this model to molecular organization, pigment density, and
Henle fiber layer geometry, subject to separate psychophysical
validation.

Soon after their discovery \cite{haidinger1844}, James Clerk Maxwell
established that Haidinger's brushes originate at the fovea of the retina
and result from selective absorption by radially oriented foveal
structures \cite{maxwell1850manuscript}.  He subsequently related the
phenomenon to the absorption of blue light by the yellow foveal pigment
\cite{maxwell1856}, an observation later known as Maxwell's spot.  It is
fitting that Maxwell's own electrodynamic equations now explain both
phenomena from first principles.


\section{Conclusions}
\label{sec:conclusions}

Starting from the complex dielectric tensor of the Henle fiber layer
\cite{misson_paper1} and applying the Berreman form of Maxwell's
equations \cite{berreman1972}, a complete Mueller--Stokes description of
the dichroism-mediated polarization percepts has been derived.  The
$4\times4$ propagation matrix
mixes the two transverse channels through a single tensor element, while
its eigenvalues stay azimuth-free, so that all azimuthal dependence is
geometric.  The resulting Mueller matrix $\bM(\vp)$ is parameterized by
four scalars $\{A,B,C,D\}$; within the intensity channel the perceived
Haidinger's brushes contrast depends only on the dichroic pair $A$ and
$B$, with the form birefringence confined to $C$ and $D$.  The
perpendicular-dipole geometry gives $B>0$ and dark arms perpendicular to
the $\mathbf{E}$-vector, with a maximum contrast $|B|/A\approx0.05$ traced
from the molecular
tilt angle.  The phenomenological model \cite{misson2003} is recovered
exactly in the perfect-dichroism limit, so that the present work supplies
its electromagnetic basis.  Previously published elaborations extend it to
a realistic pigment distribution \cite{misson2019,misson2025} and to the
Maxwell's spot and Haidinger's brushes partition \cite{misson2020}
(Supplement~1).

Because the percept is set by the local dichroic interaction at each
retinal point, the framework applies pointwise to any incident
polarization field, generating the percepts of depolarized, uniform,
spatially patterned and vector-vortex light from one relation
(Fig.~\ref{fig:percepts}).  Haidinger's brushes are the uniform-field
limit of that single macular dichroism mechanism.  Maxwell's spot is its
depolarized limit, $p=0$, and is properly a percept of macular absorption
rather than of macular dichroism: it is carried by $A$ alone, and it
matters here as the background against which the dichroic percepts, all
carried by $B$, are judged.  The azimuthal order of the vector-vortex
percept, $N=|\ell-2|$, is fixed by the topological charge alone for the
macular layer in isolation, and survives a physiological corneal
retardance, which adds a weaker $|\ell+2|$ harmonic that makes the arms
unequal without changing their number.  Percepts of
scattering origin, Boehm's brush and its spin--orbit variants among them
\cite{boehm1940,pushin2026boehm}, are not addressed by this mechanism and
require a separate treatment.  The tensor
of the companion paper together with the Mueller--Stokes forward model
developed here provide the electromagnetic foundation for the interaction
of structured and OAM-modulated light with the eye.


\begin{backmatter}
\bmsection{Funding}
D.A.P is supported by the Natural Sciences and Engineering Research Council of Canada grant [RGPIN-2024-05220] and the Canada First Research Excellence Fund


\bmsection{Disclosures}
D.A.P. and D.S. are founders of Incoherent Vision Inc. S.E.T. is founder of Azul Optics Ltd. Both companies develop devices for polarization perception. The three authors listed as founders of the two companies are the major shareholders of their companies. D.A.P. and D.S. have patents on using structured light for creating entoptic profiles. S.E.T. has patents on using polarized light for creating entoptic profiles.

\bmsection{Data availability}
No data were generated or analyzed in the presented research.  The
simulation code underlying Figs.~\ref{fig:tensor} and~\ref{fig:percepts}
and Supplement~1, Fig.~S1 is available from the corresponding author upon
reasonable request.

\bmsection{Supplemental document}
See Supplement~1 for supporting content.

\end{backmatter}

\bibliography{haidinger_v4}

\end{document}